# Shape2SAS – a web application to simulate small-angle scattering data and pair distance distributions from user-defined shapes

**Andreas Haahr Larsen[a*], Emre Brookes[b], Martin Cramer Pedersen[c] and Jacob Judas Kain Kirkensgaard[c,d]**

[a] University of Copenhagen, Department of Neuroscience, Copenhagen, Denmark

[b] University of Montana, Missoula, Montana, USA

[c] University of Copenhagen, Niels Bohr Institute, Copenhagen, Denmark

[d] University of Copenhagen, Department of Food Science, Copenhagen, Denmark

* Corresponding author: andreas.larsen@sund.ku.dk

**Abstract**

Shape2SAS is a web application that allows researchers and students to build intuition and understanding of small-angle scattering. It is available at https://somo.chem.utk.edu/shape2sas. The user defines a model of arbitrary shape by combining geometrical subunits, and Shape2SAS then calculates and displays the scattering intensity, the pair distance distribution as well as a visualization of the user-defined shape. Simulated data with realistic noise are also generated. We demonstrate how Shape2SAS can calculate and display the different scattering patterns for various geometrical shapes, such as spheres and cylinders. We also demonstrate how the effect of structure factors can be visualized. Finally, we show how multi-contrast particles can readily be generated, and how the calculated scattering may be used to validate and visualize analytical models generated in analysis software for fitting small-angle scattering data.

# Introduction

We introduce Shape2SAS, a website for simulating small-angle scattering data and pair distance distributions from various shapes. Shape2SAS is readily available as a web application (https://somo.chem.utk.edu/shape2sas) implemented using the GenApp framework [1] for constructing graphical user interfaces for scientific software. The user can construct arbitrary shapes by combining geometrical subunits such as spheres, cylinders, cubes, or ellipsoids. Each subunit is assigned an excess scattering length density (contrast), and the small-angle scattering intensity is calculated for the specified shape composed of these subunits. The website is particularly useful for educators when organizing introductory or advanced courses in small-angle scattering data analysis. Due to its nature as a stand-alone web application, the only requirement is a web browser. The program can likewise be used when developing analytical form factors, as demonstrated in this paper.

Shape2SAS applies the method previously implemented in the program McSim [2,3], and has a similar user input interface for generating and positioning shapes. The web application for McSim is, however, no longer available, and additionally, the program was written in Fortran and is not documented besides from the paper. Therefore, Shape2SAS is written in Python3 to make it possible to adjust and maintain for a broader scope of developers, and the web application is available via GenApp [1]. This allowed us to add several central features, including direct comparison of two models and inclusion of structure factors.

Shape2SAS can be used in a tutorial, where the scattering from one or two different shapes are calculated and compared. Parameters or models can be adjusted to get an intuitive feeling for the effect of these. The simulated data can then be used as output from virtual experiments used for further tutorials in (or tests of) programs for analysis of experimental small-angle scattering data, e.g., programs for fitting analytical form factors, such as SasView (https://www.sasview.org/), SASfit [4] or WillItFit [5], or programs for *ab initio* modelling [6,7].

# Applied small-angle scattering theory

In this section, we provide the theoretical and computational basis for Shape2SAS, and account for the implementation of this.

## Scattering from identical particles in solution

The scattering intensity from a diluted sample of identical particles in solution is given by the Debye equation [6], a double sum over the $N$ scatterers in each particle:

$$I(q) = n \sum_{j,k=1}^{N} \Delta b_j \Delta b_k \frac{\sin(qr_{jk})}{qr_{jk}},$$

where $n$ is the number density of the particles, $\Delta b_j$ is the excess scattering length of the $j^{\text{th}}$ scatterer, $r_{jk}$ is the distance between the $j^{\text{th}}$ and $k^{\text{th}}$ scatterer and $q$ is the momentum transfer $q = 4\pi\sin(\theta)/\lambda$, where $2\theta$ is the scattering angle and $\lambda$ is the wavelength of the incoming X-ray or neutron beam.

## The pair distance distribution

By binning the scattering pairs after their pair distances, $r$, the double sum can be reduced to a single sum over the number of bins:

$$I(q) = n \sum_{i=1}^{N_{bins}} p_i \frac{\sin(qr_i)}{qr_i},$$

where $r_i = \left(i - \frac{1}{2}\right) dr$, and $dr$ is the bin width. $p_i$ is the number of pairs in each bin, weighted by the product of their excess scattering lengths:

$$p_i = \sum_{j,k=1}^{N} f(r_{jk}) \Delta b_j \Delta b_k, \qquad \text{where } f(r_{jk}) = \begin{cases} 1, & \text{if } (i-1) \cdot dr \leq r_{jk} < i \cdot dr, \\ 0, & \text{otherwise.} \end{cases}$$

We assume point scatterers, so the effect of atomic form factors is neglected. This is also a double sum, but unlike the Debye sum, sin(x)/x is not evaluated for each distance, only for the binned histogram.

The pair distance distribution $p(r)$ is the continuous limit of $p = \{p_1, p_2, ..., p_{N_{bins}}\}$, as $N_{bins} \to \infty$ and $N \to \infty$. In this limit, the contributions from the self-terms ($i = j$) are negligible. Therefore, the self-terms are excluded from the sums in Shape2SAS, when estimating $p(r)$ and $I(q)$, and $p(0) = 0$ is added to the distribution.

The user provides the scattering length densities as input, $\Delta \rho_i = \Delta b_i / V_i$, where $V_i$ is the effective volume of the $i^{th}$ scatterer. The point density is kept constant, so $V_i$ is also constant. The calculated scattering is normalized with the forward scattering, so $I(0) = 1$. The output is therefore the unitless form factor of the particle, $P(q)$, unless an optional structure factor is selected (see section Structure factors). In this normalization step, the effective volumes of the scatterers vanish, along with the number density.

The largest distance in the molecule, $D_{max}$, is given as output, along with the radius of gyration:

$$R_g^2 = \frac{1}{2} \frac{\int_0^{D_{max}} r^2 p(r) dr}{\int_0^{D_{max}} p(r) dr}.$$

The output $p(r)$ is normalized, so the maximum is unity.

**Polydispersity**

Polydispersity in a sample means that all molecules are not identical. There can be polydispersity in size, aggregation number, in one axis or several axes, and polydispersity can be approximated by assuming a Gaussian distribution over a parameter in a model. In Shape2SAS, one general type of polydispersity is implemented:

$$p_{poly}(r) \propto \int_{1-3\sigma_{poly}}^{1+3\sigma_{poly}} p_s(r) e^{-\frac{1}{2}\left(\frac{s-1}{\sigma_{poly}}\right)^2} v^2 ds,$$

where $s$ is a scaling of all pair-distances in the user-defined shape. $p_s(r)$ is the pair distance distribution for the shape scaled with s, and $\sigma_{poly}$ is a relative polydispersity. $v = s^3$ is a relative (unitless) volume, which is included to account for the fact that the scattering from a molecule is proportional to the square of the molecular volume. The relative volume is an approximation and is only exact for spheres. The relative polydispersity, $\sigma_{poly}$, is an optional user input in Shape2SAS and by default, there is no polydispersity.

While this is an oversimplification of the many ways in which a sample can exhibit polydispersity, the implementation is simple and allows users and students to explore the consequences of polydispersity in small-angle scattering data. Additionally, it is computationally efficient compared to the alternative of generating a series of structures describing the desired polydispersity.

**Structure factors**

Interparticle interactions can be expressed in the form of structure factors, and the scattering intensity can be expressed as a product of the form factor and structure factor [4]:

$$I(q) \propto P(q) \cdot S(q).$$

So far, two structure factors are implemented in Shape2SAS: a hard-sphere structure factor [7], describing interparticle repulsion, and a 2-dimensional fractal structure factor, describing particle aggregation [8]. The decoupling approximation is also applied to account for non-spherical or polydisperse particles [8,9]. Albeit such interparticle interactions would affect the effective pair distance distribution, the program only applies the structure factor on the calculated scattering intensity. That is, if a structure factor is opted for, the output $p(r)$ is from a non-interacting molecule, whereas $I(q)$ is from a sample of interacting molecules. The $p(r)$ of the interacting molecules can, e.g., be generated in BayesApp [10], which is likewise available as a web application in GenApp.

**Interface roughness**

A composite particle can be modeled by combining geometrical subunits. If the subunits have different contrasts, there may be discrepancies between the sharp boundaries between subunits of the model and the softer and more fluent boundaries between components in the actual particle. This can be amended by including a surface roughness in the model [11,12], effectively smearing interfaces:

$$I_{rough}(q) = I(q) \cdot e^{-\frac{1}{2}(q\sigma_R)^2},$$

where $\sigma_R$ is a normal distributed smearing parameter. In Shape2SAS, surface roughness can be included as an option, and like the structure factors, it only affects $I(q)$, not $p(r)$.

**Simulated experimental noise**

Shape2SAS outputs simulated data, $I(q)$, which is generated from the calculated scattering, $I(q)$, and empirically estimated errors [13]:

$$\sigma(q) = s \cdot \sqrt{\frac{5 \cdot 10^6 + I(q)}{0.05 \text{ Å} \cdot q}},$$

where $s$ is a scaling constant, that can be changed by the user to adjust the relative noise. The constants $5 \cdot 10^6$ and 0.05 Å are chosen to imitate the noise of typical synchrotron SAXS data. The simulated data, $I_{sim}(q)$, are sampled from normal distributions with mean $I(q)$ and standard deviations $\sigma(q)$.

# Implementation

**Architecture**

The program is designed for ease of use and ease of maintenance. Input parameters are provided via an online GUI (Figure 1). Inputs are read by a Python wrapper script. The wrapper calls Python functions which perform the calculations and returns output files, plots, and values to the wrapper. Finally, the wrapper returns the output to the GUI, which displays them for the user. Thus, the program is modular, so functions can be altered, tested, expanded, or added.

**Description of core functions**

A core element of Shape2SAS is the Python function that generates points from a given subunit. First, the limits of the geometric subunit are defined in either cartesian, polar or spherical coordinates. Then, random, uniformly distributed points are generated in the volume defined by the subunit. Lastly, if selected, the points are shifted to a new center of mass. To ensure constant point density (see theory section), the volumes of the subunits are first calculated, and the number of points inserted in each subunit is adjusted, so the point density is constant. The total number of points in the model (composed of one or more subunits) is $N = 5000$. This number of points was chosen to balance precision and computational cost, which goes as $N^2$. With 5000 points, the calculated intensity is precisely reproduced up to around $q = 0.2$ Å$^{-1}$, depending on the complexity and size of the user-defined shape (Figure S1). In case of overlap, overlapping points are removed by default at this stage, but they can be included and exploited in multi-contrast situations (see Example 3).

Another core element is the calculation of all distances in a model. This step is time- and memory consuming and as such, one of the computational bottlenecks of the program. However, using the NumPy [14] function meshgrid(), all distances between 5000 points are calculated in less than a second on the GenApp server. Another computational bottleneck is the calculation of $p(r)$. This is done using the NumPy function histogram(). When polydispersity is included, several histograms are calculated, and in that case, Shape2SAS uses the histogram1d( ) function (https://github.com/astrofrog/fast-histogram) which is faster than NumPy's histogram(), but does not output $r$-values. A polydisperse $p(r)$ is thus calculated in a few seconds.

**User interface**

A key goal of the project was accessibility and ease of use. Therefore, the program is implemented as a web application, meaning that no installation is required. This is optimal for successful use in courses and tutorials. The program is part of the GenApp [1] vision for making small-angle scattering software available for everyone.

The user builds up a model of a predefined set of geometric subunits, which currently include: sphere, tri-axial ellipsoid, cylinder, disc, cube, cuboid, hollow sphere, hollow cube, cylindrical ring and discoidal ring. The geometrical parameters of the subunits can then be changed from the default values, along with their contrast and center of mass. A model consists of the collected points in the volume spanned by the subunits. By default, points are deleted from overlapping regions, but this is optional as mentioned above. If a structure factor is selected, the scattering contribution from the structure factor will be displayed along with the total scattering.

The user can choose to calculate $p(r)$ and $I(q)$ from an additional model, and the procedure is the same as for the first model. The $p(r)$ and $I(q)$ from the two models are plotted together in the GUI and can thus be directly compared without having to plot the data in third-party software. The simulated data of Model 2 is scaled by 100 by default in the plot (Figure 1E), but this scaling can be adjusted in the GUI.

All output data ($p(r)$, $I(q)$ and $I_{sim}(q)$) as well as plots and 3D model for visualization, can be downloaded from the web interface for further analysis, plotting, etc.

**Documentation and validation**

All menus and input boxes in the GUI are described with a help text, which is shown by hovering the mouse over that element. The source code is documented with extensive comments in the code, including documentation of all functions. Central references are provided directly in the GUI.

Each model is visualized as a 3D point model using a Jmol (http://www.jmol.org/) plugin and as 2D projections. The 3D model is also exported in the standard protein data bank (PDB) format, for customizable visualization and rendering.

Shape2SAS models and structure factors were tested against analytical models, using SasView (https://www.sasview.org).

Source code is available on GitHub (https://github.com/ehb54/GenApp-Shape2SAS) under the GNU General Public License v3.0.

# Examples of use

The examples are designed to be relevant in the context of research as well as research-based teaching. The first example showcases how Shape2SAS can be used for generating intuition about small-angle scattering intensity of different shapes, and the second example demonstrates the effect of inter-particle interactions. The last example demonstrates how more complex particles can be built and how Shape2SAS can be utilized to test analytical form factors.

**Example 1: Comparing scattering from different shapes**

Shape2SAS can be used to quickly calculate and compare the $p(r)$ and $I(q)$ from various geometrical bodies. One example is the scattering from a sphere with radius of 50 Å and a cylinder with the same radius and length of 400 Å (Figure 1). Such an example could help a student build intuition about the scattering and pair distance distribution functions for, e.g., spherical, elongated or hollow bodies. Moreover, the $R_g$ and $D_{max}$ are calculated and displayed in the GUI for quick comparison.

(a)

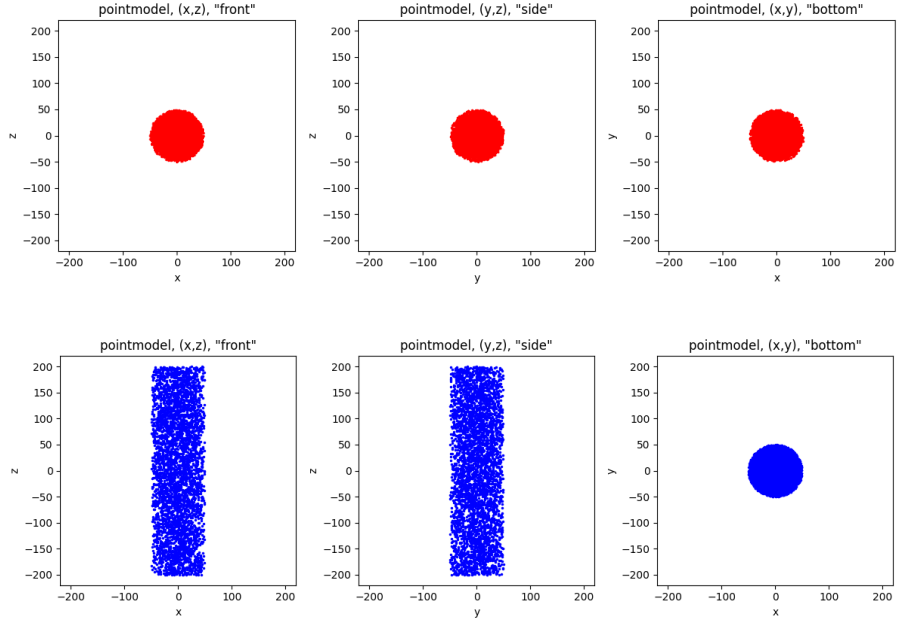

(b)

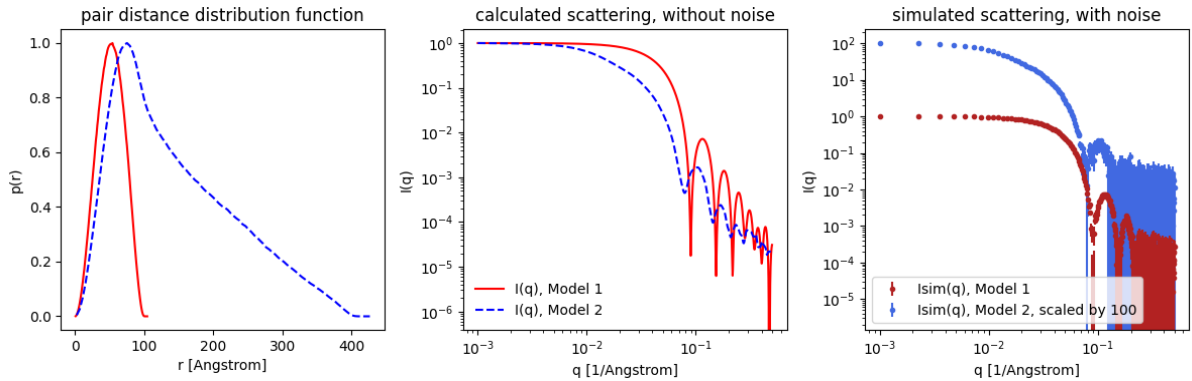

(c) (d) (e)

**Figure 1. Simulating scattering from a sphere (red) and a cylinder (blue) using Shape2SAS.** (a) GUI with input parameters. (b) 2D projections of the models. (c) Pair distance distributions. (d) Calculated scattering. (e) Simulated scattering intensities with noise. Plots are shown as they appear in Shape2SAS.

**Example 2: Hard-sphere structure factor**

Shape2SAS can add inter-particle interactions to the scattering, using built-in structure factors. One example is shown in Figure 2, where the scattering from a sample of ellipsoids of revolution (minor axis 50 Å and major axis 100 Å) was calculated with and without interparticle interaction, described by the hard-sphere structure factor and the decoupling approximation, with a hard-sphere radius of 70 Å and volume fraction of 0.2. Such an exercise could help students or researchers to understand the effect of structure factors and recognize interparticle interactions in measured small-angle scattering data.

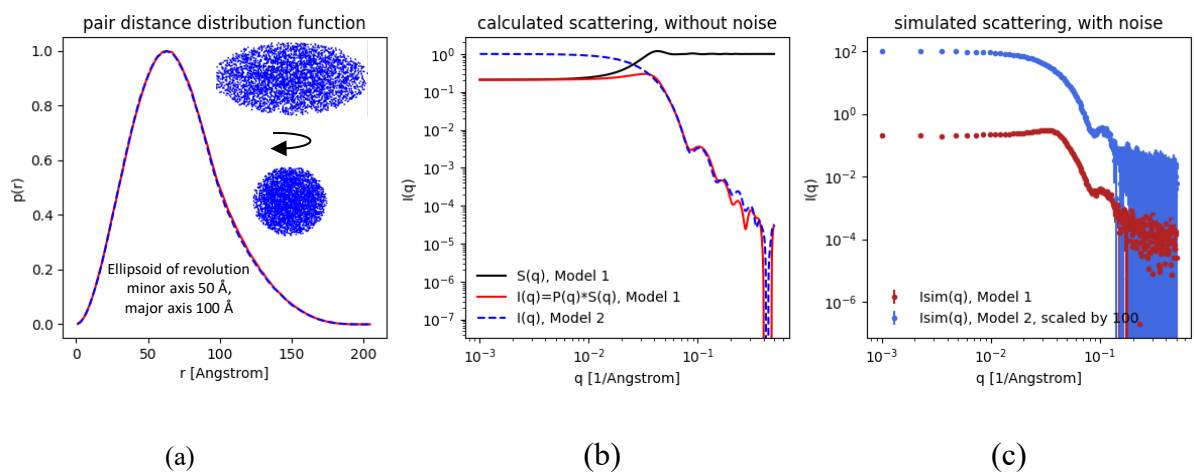

(a)            (b)            (c)

**Figure 2. Scattering intensity from ellipsoids with and without interparticle repulsion.**
(a) Pair distance distributions from two instances of an ellipsoid of revolution, without interparticle repulsion. Insert: 2D projections of the ellipsoid along the minor and major axes.
(b) Calculated scattering intensity from the ellipsoids with and without interparticle repulsion.
(c) Simulated scattering, with noise. Interparticle repulsion was described by a hard-sphere structure factor.

**Example 3: Validating analytical form factor**

Shape2SAS can be used when developing analytical form factors or deriving new ones. By generating the same shape in Shape2SAS as that of the analytical model, the simulated data from Shape2SAS can be fitted. In this example, a core-shell cylinder was simulated (core radius 20 Å, core length 360 Å, shell thickness 20 Å, core contrast -1 and shell contrast +1). The example also showcases how multi-contrast particles can be generated in two ways in Shape2SAS. In the first approach, the core-shell cylinder is generated by combining non-overlapping subunits: a cylinder core, a hollow cylindrical shell, and two small cylinders with shifted center of mass as end caps (Figure 3a). In the second approach, a large shell cylinder and a smaller core cylinder are combined, and points from the shell cylinder are removed from the overlapping region (Figure 3b). Both result in the same model, $p(r)$ and $I(q)$ (Figure 3c-d). If there is overlap, and exclusion of overlapping points is *not* opted for, then the contrast in the overlap region will effectively be the sum of contrasts of the overlapping subunits.

The simulated data were fitted with the analytical model *core_shell_cylinder* from SasView (http://www.sasview.org/sasmodels/model/core_shell_cylinder.html). Data were well-described by the model (Figure 3d) and the model parameters were refined to values consistent with the input parameters (core radius 20.10 ± 0.07 Å, core length 363 ± 2 Å, shell thickness 20.1 ± 0.1 Å, and shell contrast +0.97 ± 0.01). Errors are standard deviations. The core contrast was fixed at -1, since the combination of fitting both core contrast, shell contrasts and scaling gave high correlation between the parameters.

The demonstrated procedure can be valuable for testing new and more complex analytical models [15], and for visualizing them as bead models.

This workflow is helpful when coding analytical models for the small-angle scattering from molecules in solution. While analytical models are often ideal for implementation in a model refinement framework (such as SasView), Shape2SAS offers a simple, graphical, and intuitive manner for testing the implementation and its accuracy – in real space.

(a)

(b)

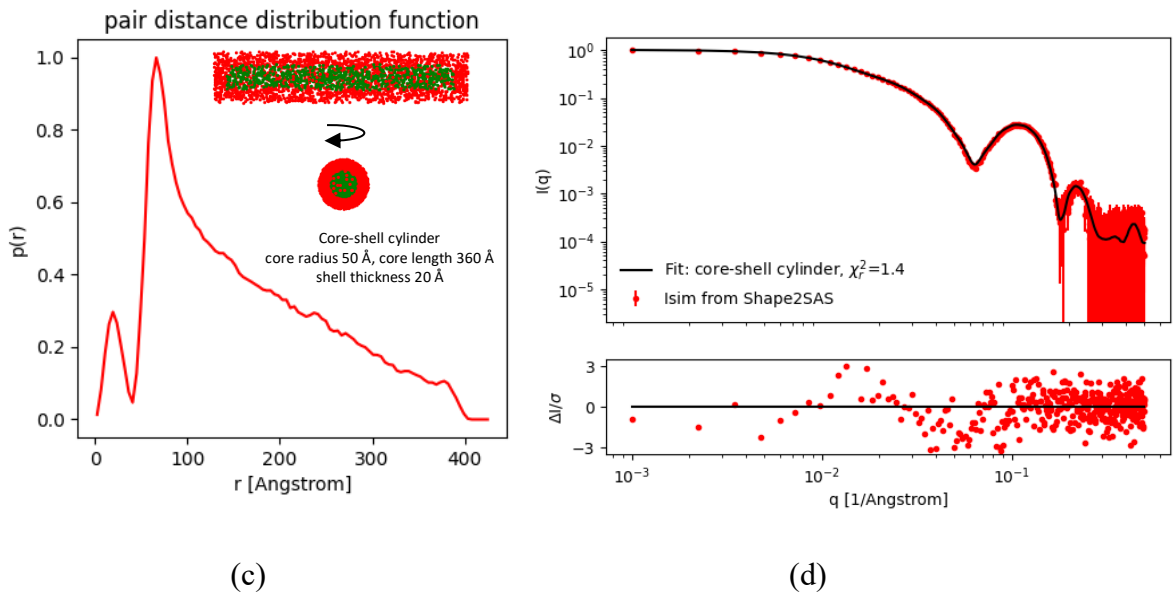

(c)                                 (d)

**Figure 3. Core-shell cylinder data, simulated in Shape2SAS and fitted with an analytical model.** (a) Core-shell cylinder input, without use of exclusion. (b) Core-shell cylinder input, using exclusion. (c) Pair distance distribution from Shape2SAS and 2D projections of the simulated beads (red: positive contrast, green: negative contrast). (d) Simulated data were fitted using the analytical model *core_shell_cylinder* from SasView.

# Discussion

Shape2SAS is useful for demonstrations and tutorials, which was showcased by three examples: Example 1 compared the small-angle scattering intensity from a sphere with the small-angle scattering intensity from a cylinder; Example 2 demonstrated the effect of interparticle repulsion via a hard-sphere structure factor; and Example 3 showcased how more advanced shapes can be built, with various excess scattering length densities and how Shape2SAS can be used to test analytical form factors, while developing these.

We note that discrepancies between the simulated scattering and an analytical model can occur from the stochastic nature of Shape2SAS (Figure 3d). Moreover, if polydispersity is fitted, this is likely implemented differently in analytical models and may result in discrepancies.

Shape2SAS is one among many programs for making virtual experiments [16], some of which can also generate 2D X-ray [17,18] or neutron [19] scattering data. Shape2SAS focuses on the data analysis step, after reduction from 2D to 1D data, and after eventual buffer subtraction. Common for such virtual experiment software is the goal of preparing the user for best use of valuable beamtime, and for helping the user to better understand and analyze the measured data.

In principle, the Shape2SAS input parameters could be transformed into variable parameters, and by addition of a scaling and a constant background, such a modified program could be used for fitting. It has previously been shown that Monte Carlo bead modeling approaches are useful in small-angle scattering analysis [20], especially for generating complicated models that are difficult to describe through analytical form factors (e.g. perforated vesicles [21] or protein-lipid complexes [22]). However, other programs, applying the same principles have been developed with fitting in mind, including CDEF [23] and SPONGE (https://github.com/bamresearch/sponge) and we recommend use of these for fitting. CDEF applies the same principle of binning scattering pairs as Shape2SAS, whereas SPONGE applies the Debye formula directly, making it more accurate at the cost of longer computational times. When fitting actual data, we moreover encourage parametrizing the model to reflect physical properties rather than geometrical, to better be able to constrain and validate the refined parameters, e.g., with biophysical assays (see, e.g. [11,24]). This is not possible in Shape2SAS and would typically have to be adjusted from case to case, which is easier without a GUI. Shape2SAS may, however, be downloaded, and thanks to the modular architecture, functions can readily be reused by other users, to accommodate specific needs.

In summary, Shape2SAS makes small-angle scattering theory intuitive, visual, playful and accessible for new and experienced users.

# Acknowledgements

This work used the Extreme Science and Engineering Discovery Environment (XSEDE) [25], which is supported by National Science Foundation Grant Number ACI-1548562 and utilized Jetstream2 [26] at Indiana University through allocation TG-MCB17057 to EB. This work benefited from CCP-SAS [27] software developed through a joint EPSRC (EP/K039121/1) and NSF (CHE-1265821) grant. AHL was funded by the Lundbeck Foundation grant R347-2020-2339. EB was funded by the National Institutes of Health, National Institute of General Medical Sciences grant GM120600 and the National Science Foundation grant 1912444. The authors would like to acknowledge Steen Hansen for valuable discussions and inspiration and the students at various courses on scattering techniques at University of Copenhagen, who have tested the program.

# Supplementary figures

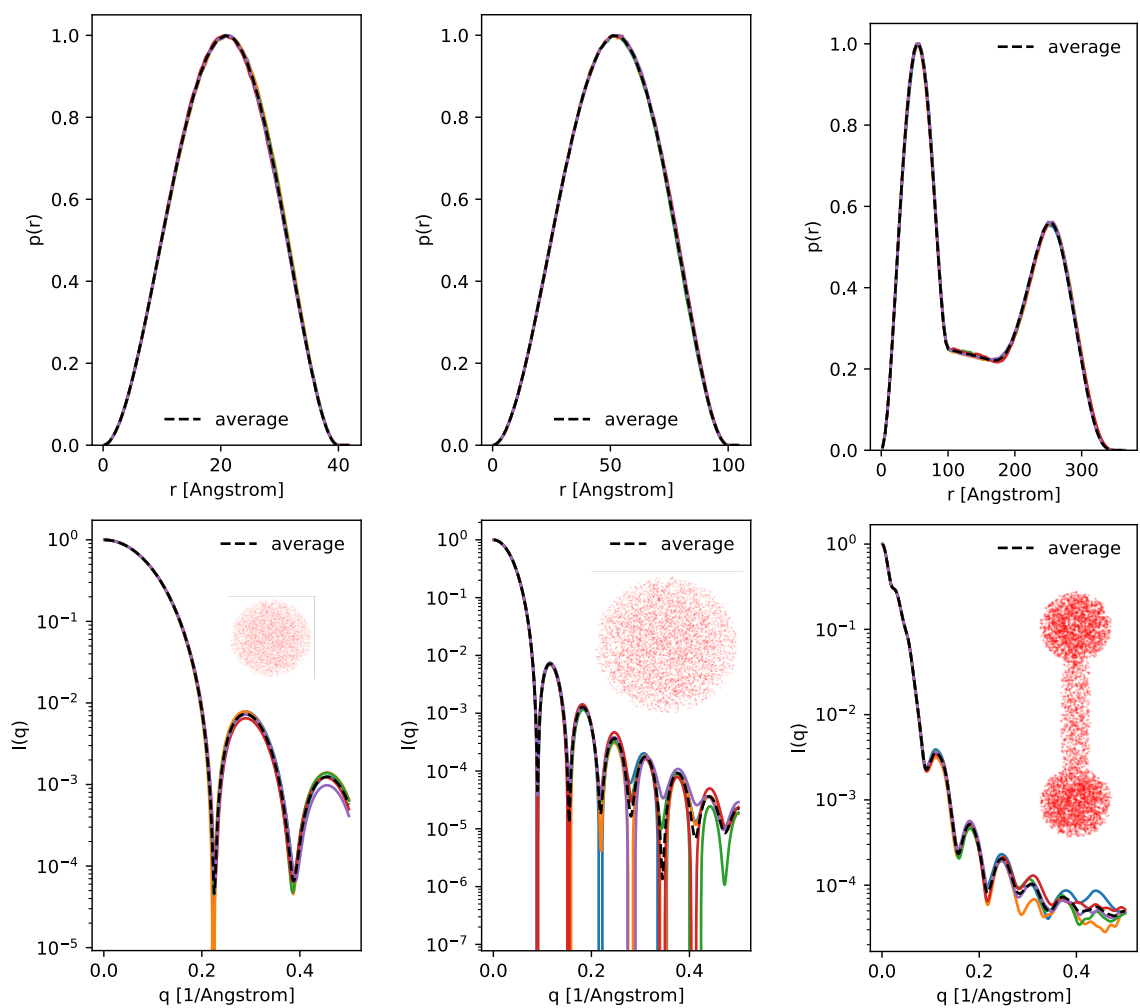

**Figure S1. Precision of Shape2SAS when using 5000 points.** Calculated pair distance distribution and intensity for five repeated simulations. (a) Sphere with radius 20 Å (b) Sphere with radius 50 Å. (c) Dumbbell composed of a cylinder with radius 50 Å and length 200 Å, and two spheres of radius 20 Å, with their center of mass shifted ±125 Å.